\documentclass[10pt,conference]{IEEEtran}
\IEEEoverridecommandlockouts
\usepackage{times}
\usepackage{tikz}
\usetikzlibrary{positioning}
\usepackage{pgfplotstable}
\pgfplotsset{compat=1.3}
\usepgfplotslibrary{fillbetween}
\usetikzlibrary{patterns}
\usepackage{subfiles}
\usepackage{xcolor}
\usepackage{cite}
\usepackage{amsmath,amssymb,amsfonts}
\usepackage{algorithmic}
\usepackage{graphicx}
\usepackage{textcomp}
\usepackage{xcolor}
\usepackage{booktabs}
\usepackage{RobStd}
\pgfmathdeclarefunction{gauss}{2}{%
  \pgfmathparse{1/(#2*sqrt(2*pi))*exp(-((x-#1)^2)/(2*#2^2))*0.7}%
}

\def\BibTeX{{\rm B\kern-.05em{\sc i\kern-.025em b}\kern-.08em
    T\kern-.1667em\lower.7ex\hbox{E}\kern-.125emX}}
\pgfplotsset{compat=1.15}
\begin{document}

\title{In-Memory Nearest Neighbor Search with FeFET Multi-Bit Content-Addressable Memories}

\author{ Arman Kazemi$^{*\star\psi}$, Mohammad Mehdi Sharifi$^{*\star\infty}$, Ann Franchesca Laguna$^{*}$, Franz M{\"u}ller$^\dagger$, Ramin Rajaei$^{*}$,\\Ricardo Olivo$^\dagger$, Thomas K{\"a}mpfe$^\dagger$, Michael Niemier$^{*}$, X. Sharon Hu$^{*}$, \{$^{\psi}$akazemi,$^{\infty}$msharif1\}@nd.edu\\
\normalsize $^{*}$University of Notre Dame, IN, US, \normalsize $^\dagger$Fraunhofer IPMS-CNT, Dresden, Germany, $^{\star}$equal contribution 
}

\maketitle

\begin{abstract}

Nearest neighbor (NN) search is an essential operation in many applications, such as one/few-shot learning and image classification. As such, fast and low-energy hardware support for accurate NN search is highly desirable. Ternary content-addressable memories (TCAMs) have been proposed to accelerate NN search for few-shot learning tasks by implementing $L_\infty$ and Hamming distance metrics, but they cannot achieve software-comparable accuracies. This paper proposes a novel distance function that can be natively evaluated with multi-bit content-addressable memories (MCAMs) based on ferroelectric FETs (FeFETs) to perform a single-step, in-memory NN search. Moreover, this approach achieves accuracies comparable to floating-point precision implementations in software for NN classification and one/few-shot learning tasks. As an example, the proposed method achieves a 98.34\% accuracy for a 5-way, 5-shot classification task for the Omniglot dataset (only 0.8\% lower than software-based implementations) with a 3-bit MCAM. This represents a 13\% accuracy improvement over state-of-the-art TCAM-based implementations at iso-energy and iso-delay. The presented distance function is resilient to the effects of FeFET device-to-device variations. Furthermore, this work experimentally demonstrates a 2-bit implementation of FeFET MCAM using AND arrays from \mbox{GLOBALFOUNDRIES} to further validate proof of concept.



\end{abstract}


\section{Introduction}
\label{sec:introduction}

Nearest neighbor (NN) search computations are at the core of many applications such as hyperdimensional computing~\cite{imani2019searchd}, classification~\cite{cover1967nearest}, and memory-augmented neural networks (MANN)~\cite{ni2019ferroelectric,laguna2019design}. NN search is defined as finding the closest point among $p$ data points to a query point in an $N-$dimensional space~\cite{bohm2001searching} based on a distance function $D$. Some commonly used distance functions include cosine, $L_\infty$, $L_2$, etc. The complexity of NN search increases with respect to $N$ and $p$, which is known as the curse of dimensionality~\cite{bohm2001searching}. This results in poor scaling of NN search in conventional hardware, and incurs high energy and latency overheads.

To address the curse of dimensionality, hardware solutions have been proposed to increase the efficiency of the NN search operations~\cite{kim201386,ni2019ferroelectric,laguna2019design}. Among these, a promising approach is to use ternary content-addressable memories (TCAMs) to accelerate NN search. TCAMs are associative memories that compare the query with stored data in parallel and return the address of the matching data~\cite{pagiamtzis2006content}. TCAMs employ a ``don't care'' state, in addition to ``0'' and ``1'' states, for wildcard operations that match both ``0'' and ``1''. Work in~\cite{laguna2019design} used multiple TCAM look-ups to implement the $L_\infty$ distance function and offered speed and latency improvements compared to GPUs. The work in~\cite{ni2019ferroelectric} demonstrated an in-memory Hamming distance measurement with CAMs and used ferroelectric field-effect transistors (FeFETs) for additional improvements due to the FeFET's compact device structure.

Both~\cite{ni2019ferroelectric} and~\cite{laguna2019design} evaluated the efficacy of CAM-based distance metrics by accelerating few-shot learning tasks. However, due to the $L_\infty$ distance function implemented in~\cite{laguna2019design}, it suffered significant accuracy loss. \cite{ni2019ferroelectric} must employ a locality-sensitive hashing (LSH) function~\cite{andoni2006near} to encode features to a binary representation in order to perform NN search with Hamming distance in TCAMs. Moreover, while energy-delay product improvements of $20\times$ are possible, classification accuracies still fall short of what is possible with software. 

Recently, analog content-addressable memories (ACAMs) have been proposed based on Resistive RAMs (RRAMs)~\cite{li2020analog} and FeFETs~\cite{yin2020fecam} which offer higher densities compared to TCAMs for exact match search. The results of the FeFET-based design are based on simulations only, while the RRAM design is demonstrated experimentally. Moreover, neither design has been used to implement a useful distance function for NN search in the literature.

In this work, we introduce the concept of multi-bit content-addressable memories (MCAMs). Specifically, we make the following contributions: (i) We propose a novel distance function that can be natively evaluated with FeFET MCAMs to accomplish single-step in-memory NN search. (ii) We apply this distance function to one/few shot learning and NN classification and achieve accuracies comparable to software implementations. As an example, we achieve a 98.34\% accuracy for a 5-way, 5-shot classification task with a MANN for the Omniglot dataset~\cite{lake2015human} (only 0.8\% lower than software-based implementations) when using a 3-bit MCAM for NN search; this is a 13\% improvement over previous TCAM-based implementations with the same-length CAM words. (iii) We study the effects of FeFET threshold voltage variations of the accuracy of the proposed distance function. (iv) We experimentally demonstrate a 2-bit implementation of the FeFET MCAM to further demonstrate the viability of this hardware. 


\section{Background and Related Work}
\label{sec:bg}
Here we define the MCAM concept, review the FeFET device~\cite{jerry2018ferroelectric}, and discuss previous FeFET CAM designs.

\begin{figure}[ht]
    \centerline{\includegraphics[width=1\columnwidth]{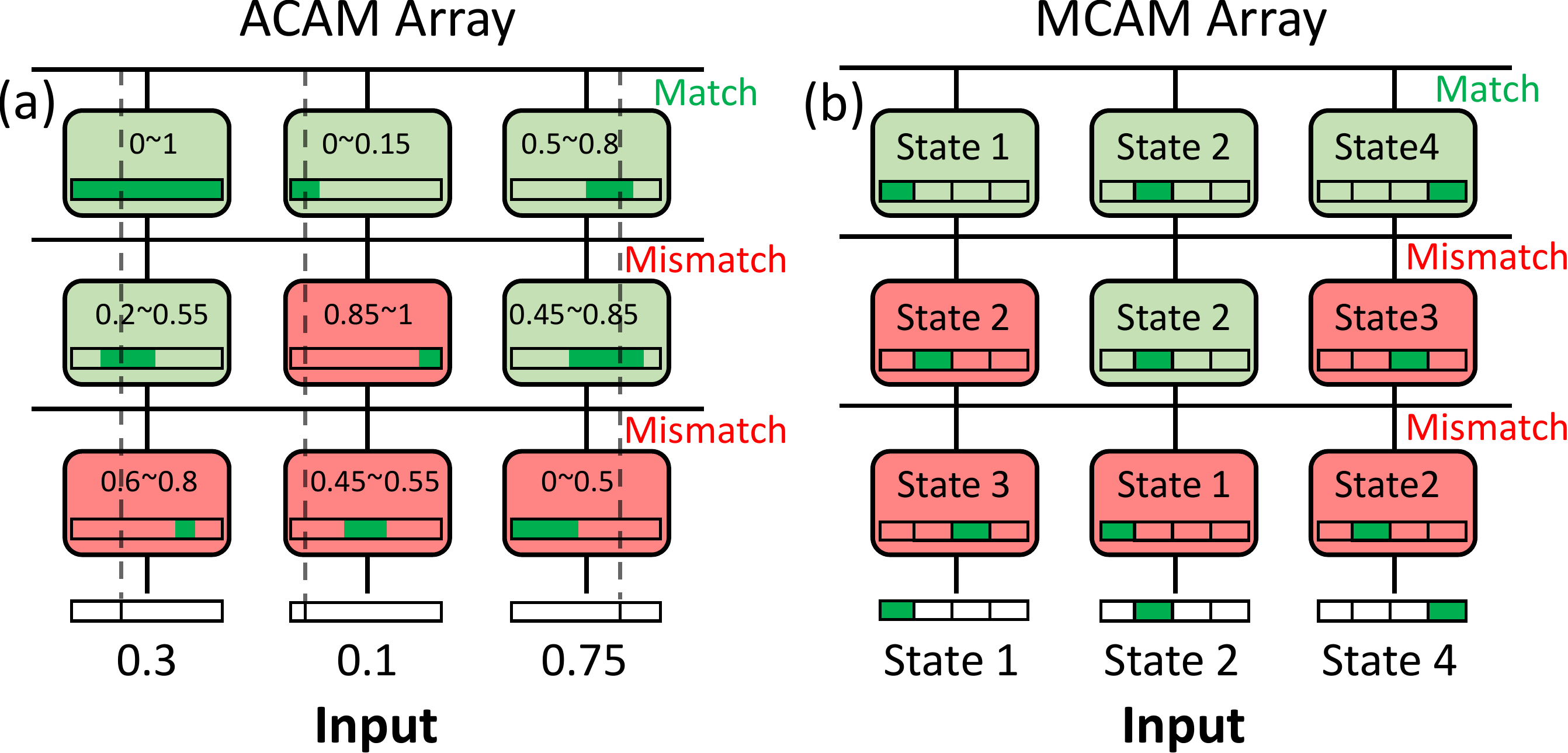}}
    \caption{Schematics for (a) Analog and (b) Multi-bit CAM concepts.}
    \label{fig:MCAM_Vs_ACAM}
    \vspace{-0.2in}
\end{figure}

\subsection{MCAM Concept}
\label{sec:bg_mcam}
MCAMs are a specific case of analog content-addressable memories (ACAMs) that were recently discussed in~\cite{li2020analog}. An ACAM cell stores a range of values and compares the analog input with the stored range to determine a match or a mismatch. In the example in Fig.~\ref{fig:MCAM_Vs_ACAM}(a), the inputs of the ACAM can be any value in the range (0,1), and the cells can store any subset of the range (0,1). Given 0.3, 0.1, and 0.75 as inputs, the top row of the ACAM array in Fig.~\ref{fig:MCAM_Vs_ACAM}(a) reports a match as its three cells store ranges (0,1), (0,0.15), and (0.5,0.8).

As mentioned in~\cite{li2020analog}, if an ACAM cell stores \textit{narrow, non-overlapping ranges}, it can be a high-density digital CAM, or as referred to here, an MCAM. In other words, every narrow range represents a state, which is depicted in Fig.~\ref{fig:MCAM_Vs_ACAM}(b). The key difference between MCAMs and ACAMs is that each MCAM cell only searches with a \textit{limited set of input values}, with each corresponding to a state (Fig.~\ref{fig:MCAM_Vs_ACAM}(b)); ACAMs, on the other hand, search with an infinite number of inputs. In MCAMs, the stored ranges have a one-to-one correlation with the inputs. Thus, if the number of \textit{narrow, non-overlapping ranges} and \textit{specific inputs} are 4, the MCAM would be a 4-state or 2-bit MCAM, as shown in Fig.~\ref{fig:MCAM_Vs_ACAM}(b). As such, MCAM is a special, highly robust case of ACAM.

\subsection{The FeFET Device}
\label{sec:bg_fefet}
\begin{figure}[t]
    \centerline{\includegraphics[width=0.8\columnwidth]{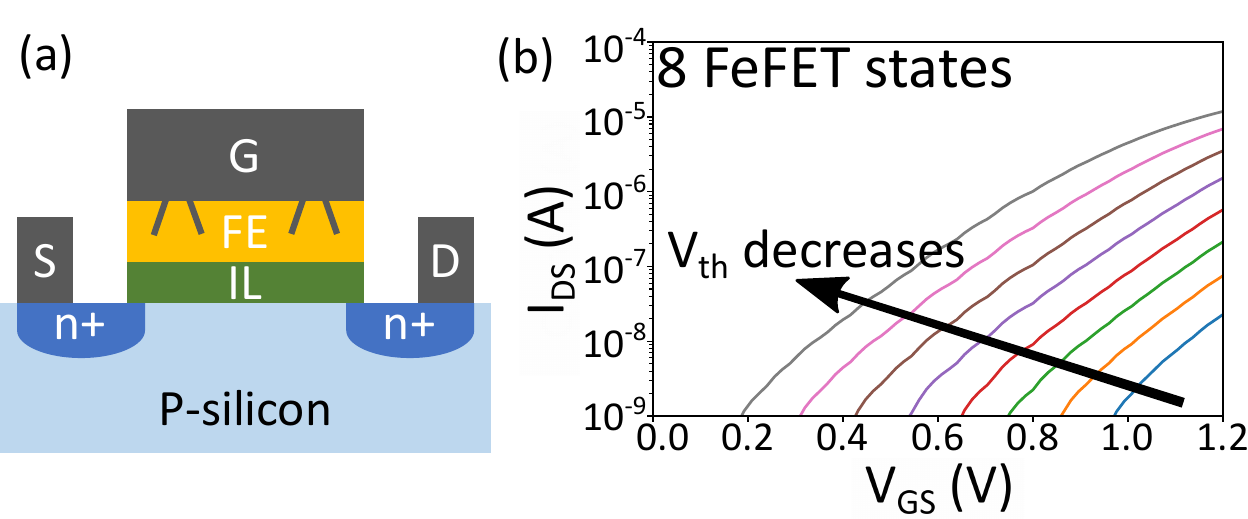}}
    \caption{(a) FeFET device~\cite{ni2018circuit}; (b) Transfer characteristics of a FeFET device programmed with different voltage pulses.}
    \label{fig:FeFET_device}
    \vspace{-0.2in}
\end{figure}

The FeFET device (Fig.~\ref{fig:FeFET_device}(a)) incorporates a ferroelectric (Fe) layer in the gate stack of a MOSFET~\cite{jerry2018ferroelectric}. The polarization of the Fe layer determines the threshold voltage ($V_{th}$) of the device. The $V_{th}$ of FeFETs can be controlled by applying voltage pulses to the gate of the FeFET. FeFETs can store multiple $V_{th}$ levels through partial polarization switching of the Fe layer. Recent works exploit the multi-level behavior of FeFETs~\cite{kazemi2020hybrid,jerry2018ferroelectric,yin2020fecam} to increase density and reduce energy and delay of their designs. We also use FeFETs in a multi-level manner in our work and use the Preisach model presented in~\cite{ni2018circuit} to model the behavior of FeFETs.

There have been multiple programming schemes proposed in the literature~\cite{ni2018circuit,jerry2018ferroelectric} to reach the intermediate states of FeFETs. In this work, we use single, same-width pulses with different amplitudes for programming. Using the Presiach model, we achieve 8 different $V_{th}$ levels for FeFETs, shown in Fig.~\ref{fig:FeFET_device}(b). However, the Presiach model does not account for the stochastic behavior of FeFET polarization switching in different FeFET devices. Thus, to be able to properly model device-to-device variation, we use a model based on a Monte Carlo framework~\cite{ni2020FeFETD2D}. This allows us to properly study the effects of device-to-device variation on the proposed distance function. Note that the model based on a Monte Carlo framework~\cite{ni2020FeFETD2D} does not have a Verilog-A implementation and we cannot use it for our SPICE simulations.

\subsection{FeFET-Based CAM Designs}
\label{sec:bg_previous_mcam}
The CAM cell in Fig.~\ref{fig:mcam_schematic_and_vbs}(a) was first proposed as a TCAM in~\cite{ni2019ferroelectric} where it can store ``0", ``1", or ``X'' with two FeFET polarization states and search with high/low inputs. A more recent work~\cite{yin2020fecam} used the same cell as an ACAM exploiting the partial polarization switching of FeFETs to increase the density of the cell. To store and search analog data in the cell, the ACAM cell needs to perform ``analog" inversions with respect to a ``\textit{center}" for both FeFET polarizations and the search inputs. The analog inverse of a signal has the same distance from the \textit{center} as the original input. For example, in Fig.~\ref{fig:mcam_schematic_and_vbs}(b), $\overline{inp1}$ and $\overline{inp2}$ are the analog inverse of $inp1$ and $inp2$. The same analog inversion applies to the $V_{th}$ level of FeFETs. An example of this is $\overline{V_{th-Lo}}$ for $V_{th-Lo}$.

Realizing an ACAM cell has two challenges: (i) The FeFETs must be programmed in a truly analog manner which is unrealistic. Only an asymmetric and nonlinear 5-bit precision (highest in literature) is achieved~\cite{jerry2018ferroelectric} with complicated programming schemes; (ii) based on our simulations (not discussed in detail here due to space limitations), a single on-the-fly analog inversion can cost $100\times$ more energy compared to an array search. Work in~\cite{yin2020fecam} assumes such analog inverter exists but does not provide design details. 

\section{The Proposed MCAM-Based Distance Function}
\label{sec:proposed}
Here we review the FeFET MCAM cell from~\cite{yin2020fecam} and explain the multi-bit input/programming scheme and its benefits. We elaborate on NN search with the described MCAM cell and discuss the proposed novel distance function for NN search.

\subsection{MCAM Cell}
\label{sec:mcam}
\begin{figure}[t]
  \centering
  \includegraphics[width=\columnwidth]{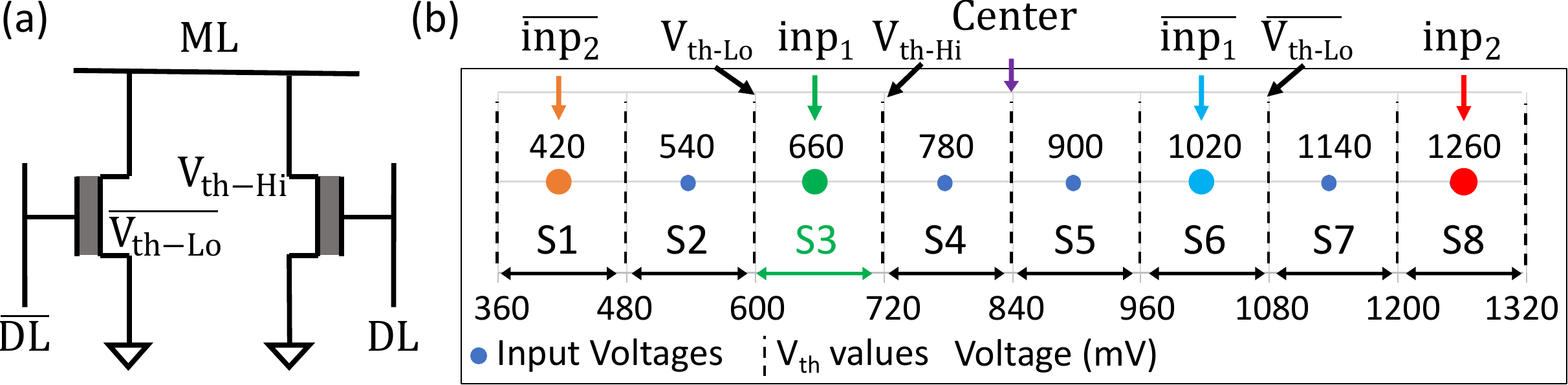}
  \caption{(a) Schematic of the considered cell from~\cite{ni2019ferroelectric,yin2020fecam}; (b) The states of the 3-bit MCAM cell denoted with S1 to S8, its 8 inputs in their corresponding states, and example analog inverse signals of inputs and $V_{th}$ values.}
  \label{fig:mcam_schematic_and_vbs}
  \vspace{-0.2in}
\end{figure}

We first discuss the search operation, schematic, and input/programming scheme of the considered MCAM cell. For the search operation with the MCAM cell in Fig.~\ref{fig:mcam_schematic_and_vbs}, the input is applied to data line ($DL$) and its analog inverse is applied to $\overline{DL}$. If the input matches the state stored in the MCAM cell, the ML stays high; otherwise, it discharges to the ground. For a conventional exact match search in a CAM array, the ML is latched after the inputs are applied to the cells of the array, and the ML that is high suggests a match between the stored row and the input. For a NN search, the inputs are applied in the same way to the cells, although there are still some differences, as will be described in Sec.~\ref{sec:distance}.

To overcome the challenges of realizing the ACAM discussed in Sec.~\ref{sec:bg_previous_mcam}, we use a multi-bit scheme that overcomes both challenges by: (i) limiting the number of states to $2^B$, where $B$ is the number of bits the cell can store ($B=3$ in Fig.~\ref{fig:mcam_schematic_and_vbs}(b)), which allows for a realistic realization of MCAMs as FeFETs can be programmed to 8 distinct polarization states as shown in Fig.~\ref{fig:FeFET_device}(b); (ii) limiting the number of search inputs to $2^B$. Note that the collection of the input signals are the same as the collection of their inverse signals. The same applies to the $V_{th}$ and $\overline{V_{th}}$ values. This means that the MCAM design \textit{does not} need to perform analog inversions on-the-fly needs and \textit{only} to generate 8 distinct programming and input voltages for a 3-bit cell.

Fig.~\ref{fig:mcam_schematic_and_vbs}(b) illustrates the states and search inputs of the MCAM cell. The cell can store any of the states S1 to S8 by programming the right FeFET in Fig.~\ref{fig:mcam_schematic_and_vbs}(a) to the $V_{th}$ on the right side of the state arrow (Fig.~\ref{fig:mcam_schematic_and_vbs}(b)) and the left FeFET to the analog inverse of $V_{th}$ to the left of the state arrow. As an example, to store state 3, the $V_{th}$ of the right and left FeFETs are programmed to $720mV$ ($V_{th-Hi}$) and $\overline{800}mV = 1080mV$ ($\overline{V_{th-Lo}}$), respectively. The search inputs are also depicted with blue dots and have a one-to-one relation with the states of the MCAM cell. Furthermore, given FeFETs that can achieve only 4 polarization states, it is straightforward to design a 2-bit MCAM cell. To this end, it suffices to combine the two neighboring states in Fig.~\ref{fig:mcam_schematic_and_vbs}(b) and have the 4 inputs in the middle of the new states.

\subsection{In MCAM Distance Function}
\label{sec:distance}

\begin{figure}[t]
  \centering
  \includegraphics[width=\linewidth]{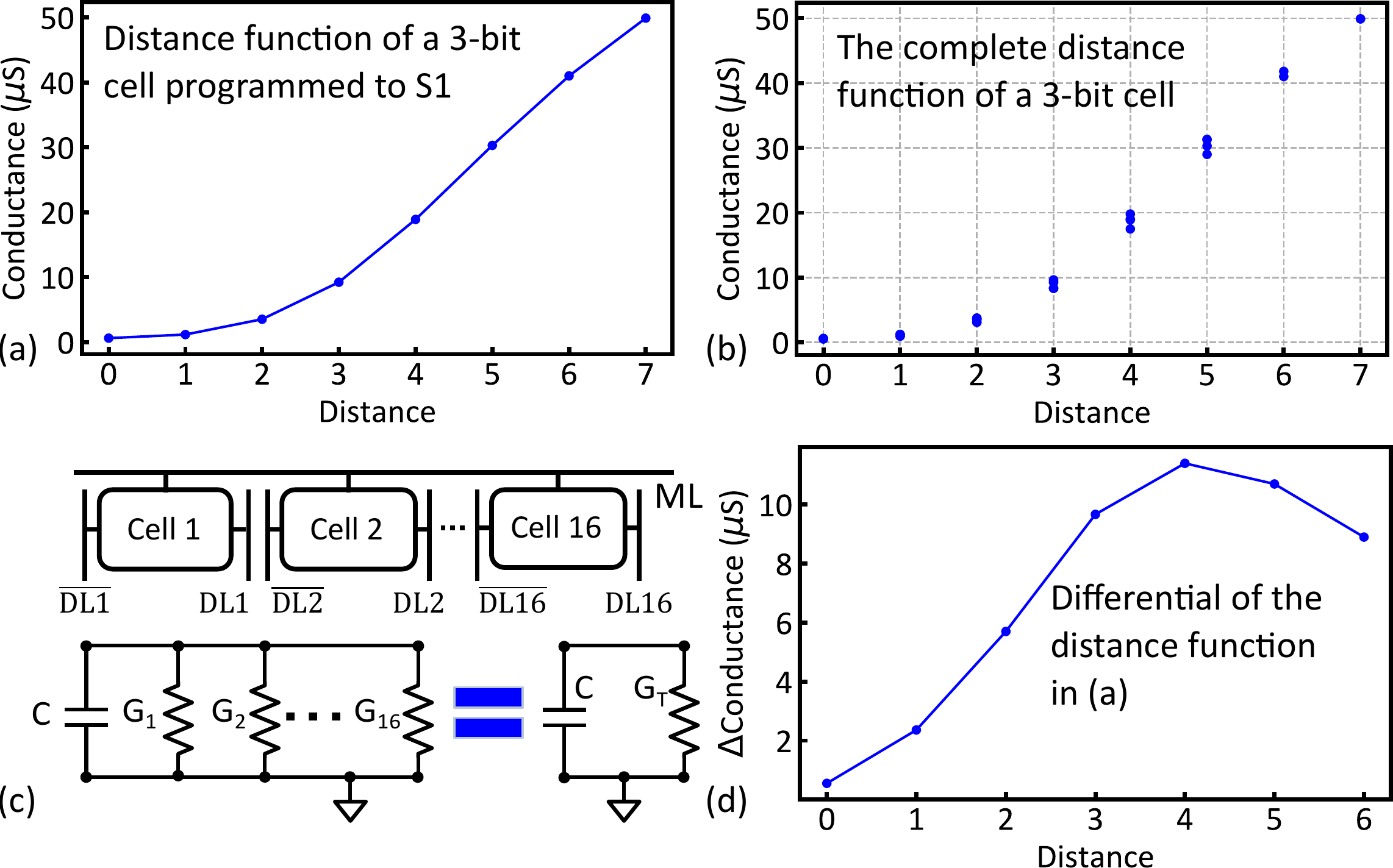}
  \caption{(a) Distance function for a 3-bit cell programmed to S1; (b) Distance function for a 3-bit cell; (c) ML discharge modelled by an RC model where the total conductance of ML is the summation of the conductance of all the cells; (d) The bell-shaped curve for the derivative of the distance function.}
  \label{fig:distance}
  \vspace{-0.25in}
\end{figure}

The distance function employed in a NN search has a significant impact on the application-level accuracy of a classification task. For an MCAM cell in state $S$ with input $I$, the distance between $I$ and $S$ is defined as $|I-S|$. Fig.~\ref{fig:distance}(a) is the conductance vs. distance curve for a 3-bit cell storing S1. This curve illustrates the distance function of the cell while storing S1 where the larger the conductance, the greater the distance. This distance function follows the transfer characteristics of FeFETs since only one of the FeFETs is contributing to the conductance of the cell. Thus, the conductance increases exponentially with respect to distance.

As Fig.~\ref{fig:distance}(a) is the distance function for a cell storing S1, a complete representation of the cell's distance function is $F(I,S) = G$, where $I$ and $S$ are the input state and the cell state, respectively. Fig.~\ref{fig:distance}(b) shows the distance function of the cell, where the different dots with the same distance (on the x-axis) are from different $I$ and $S$ pairs. The differences in the conductance of the different instances of a distance are due to the variations of the transfer characteristic of the FeFETs in different states captured by the Presiach model~\cite{ni2018circuit}. Same as Fig.~\ref{fig:distance}(a), the distance function follows the exponential transfer characteristics of FeFETs as with any combination of input and cell states only one of the FeFETs is ``On' and the other is ``Off". The proposed distance function achieves high application-level accuracies as we will show in Sec.~\ref{sec:eval}.

We investigate the ML discharge process of the MCAM to better comprehend the parameters that reflect the distance between a query and the stored MCAM entries. As illustrated in Fig.~\ref{fig:distance}(c), the ML discharge in the MCAM can be modeled by an RC model. The ML is pre-charged to 0.8V. The capacitance of all the rows (C in Fig.~\ref{fig:distance}(c)) are fixed and the same, as all devices are the same for each row. Each cell has a fixed conductance $G_i$ based on its cell state and input state. Then, the conductance of a row is just the addition of the conductances of all its cells ($G_T = G_1 + G_2 + \dots + G_{16}$). As such, $G_T$ directly reflects the distance between the input vector and the memory where a higher $G_T$ represents a greater distance. Although measuring $G_T$ directly is not feasible, it is possible to identify the ML for which the voltage discharges the slowest; that ML has the shortest distance from the input. The sense amplifier presented in~\cite{imani2019searchd} can be used to this effect. This sense amplifier is analyzed in detail in~\cite{imani2019searchd}. 

To understand the effects of the exponential increase of cell conductance with respect to the distance in measuring the distance in an array, we define $G_{n\_d}$ as the conductance of an MCAM row when all of its cells observe distance 0, except for $n$ cells which observe distance $d$. The total distance of a row with conductance $G_{n\_d}$ is $n\times d$. Simulations of a 16-cell, 3-bit MCAM row suggest that $G_{1\_4} > G_{4\_1}$, even though both rows have a total distance of 4. Similarly, $G_{1\_7} \gg G_{7\_1}$ further stresses the significance of the exponential relation between distance and cell conductance (Fig.~\ref{fig:distance}(b)). Unsurprisingly, $G_{1\_4} > G_{7\_1}$ suggests that rows with lower distance mismatches that are concentrated in a single cell have higher conductance than rows with higher distance mismatches with the distance spread among different cells.

To analyze the effectiveness of the proposed distance function for NN search applications, we consider the derivative of the distance function in Fig.~\ref{fig:distance}(a) with respect to distance (shown in Fig.~\ref{fig:distance}(d)). The derivative of the conductance is the highest when the cell observes 3 to 5 distances as opposed to when the points are near each other (0 to 2 distances away). Moreover, when the points are too far from each other (6-7 distance away), there is a drop in the derivative, which shows lesser significance for points that are already too far. This behavior is suitable for NN Search, as the distance function differentiates the most between the points that are in the gray area of neither near nor far. Results in Sec.~\ref{sec:eval} will corroborate our analysis. To the best of our knowledge, the proposed distance function has neither been used for NN search in software nor been derived from a circuit.

\subsection{Effects of FeFET Threshold Voltage Variations}
\label{sec:vth_var}
\begin{figure}[t]
  \centering
  \includegraphics[width=0.95\linewidth]{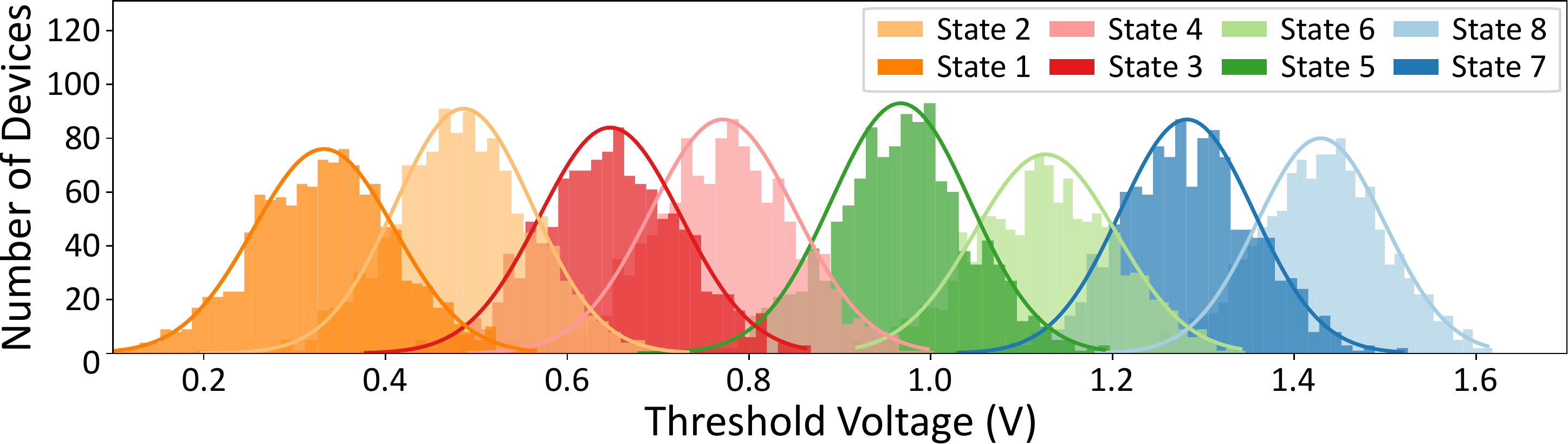}
  \caption{$V_{th}$ distribution of 1200 FeFET devices based on the model from~\cite{ni2020FeFETD2D} with sigma of variations up to 80mV.}
  \label{fig:FeFET_vth_variation}
  \vspace{-0.2in}
\end{figure}

FeFET $V_{th}$ variations directly affect the behavior of the proposed distance function by changing the conductance of the MCAM cell for different $I$ and $S$ pairs. This could affect the accuracy of NN search with MCAMs for different applications. Variations in FeFET domain switching~\cite{ni2020FeFETD2D} affect the $V_{th}$ of FeFETs. Thus, we consider 1200 devices with a channel length and width of 250nm and 250nm for variation studies and simulate them with the model from~\cite{ni2020FeFETD2D}. We program each device to 8 states with single, same-width pulses with different amplitudes (no verification pulses), as in Sec.~\ref{sec:mcam}. Fig.~\ref{fig:FeFET_vth_variation} shows the distribution of FeFET $V_{th}$ values for the 8 states. We model these variations as Gaussians for evaluations in Sec.~\ref{sec:few-shot}.

\section{Evaluations and Results}
\label{sec:eval}

Although CAM-based NN search operations are extremely efficient, they often struggle to achieve state-of-the-art accuracies at the application level when compared to software implementations. Here, we describe and compare three state-of-the-art NN search implementations: (i) software running on GPUs, (ii) a TCAM-based solution~\cite{ni2019ferroelectric}, and (iii) the proposed distance function using the FeFET MCAM. To evaluate the application-level accuracy of the proposed distance function, we consider NN classification and one/few-shot learning with MANNs. Furthermore, we analyze the effects of device-to-device variation on the deliverable accuracy of one-few shot learning based on our simulations in Sec.~\ref{sec:vth_var}. Finally, we present a preliminary experimental demonstration of a 2-bit FeFET MCAM on a FeFET AND array manufactured by \mbox{GLOBALFOUNDRIES} and evaluate its behavior.

\subsection{NN Search Implementations}

Here we describe the three implementations considered in this paper. The first is a GPU implementation of NN search between a query and data points stored in the main memory on a GPU. Different distance functions are used to calculate the similarity of the real-valued query with the real-valued memory entries in single precision floating-point (FP32) form. We consider the cosine and Euclidean distance functions, as they achieve state-of-the-art accuracy for the intended applications~\cite{snell2017prototypical}. Such distance calculations require memory transactions to read memory entries, which can be expensive from the perspective of both time and energy~\cite{han2016eie}.

The second implementation~\cite{ni2019ferroelectric}, leverages Hamming distance measurements in TCAMs. As a Hamming distance measurement cannot achieve competitive application-level accuracies, all the features of the real-valued query and memory entries are transformed using an LSH algorithm~\cite{andoni2006near} run on a GPU to create intermediate binary signatures. A TCAM array stores the LSH signatures of the memory entries in its rows (a one-time programming overhead) and measures the Hamming distance of the LSH signature of the query with the TCAM rows. The TCAM row with the shortest Hamming distance from the query is the nearest neighbor. As Hamming distance measurements happen for all rows at once, and data transfer operations are not required, the TCAM+LSH approach is more energy and time-efficient than the GPU-based approach\cite{ni2019ferroelectric}. Although, this efficiency comes at the cost of accuracy loss.

Third, to perform NN search with the FeFET MCAM, the real-valued features of the query and memory entries are quantized to the same bit precision as the MCAM, e.g., 3 bits. The quantized features of the memory entries and the query map to MCAM cells (a one-time programming overhead) and inputs of the MCAM cells, respectively. Since the features are quantized to the same bit precision as the MCAM, there is a one-to-one mapping for memory entry and query features to MCAM cell states and input states. The MCAM performs a single-step, in-memory, NN search based on the proposed distance function to find the nearest neighbor of the query.

To properly simulate the behavior of the proposed distance function $F(I,S)=G$ (refer to Sec.~\ref{sec:distance}), we create a 2D conductance look-up table based on states and inputs for a single cell and store it in a Python array. The run-time conductance of each cell is read from the look-up table based on the state of the stored feature and the input feature. For example, for a 3-bit cell storing S5 and an input in S3, the conductance is pulled from the look-up table entry (5,3). For each MCAM row representing a data point, the conductances of all the cells (each cell representing a feature) are summed up to get the total conductance of that row. The MCAM row with the lowest conductance is chosen as the nearest neighbor of the query as discussed in Sec.~\ref{sec:distance}.

\subsection{NN Classification}

\begin{figure}[t]
  \centering
  \includegraphics[width=0.95\linewidth]{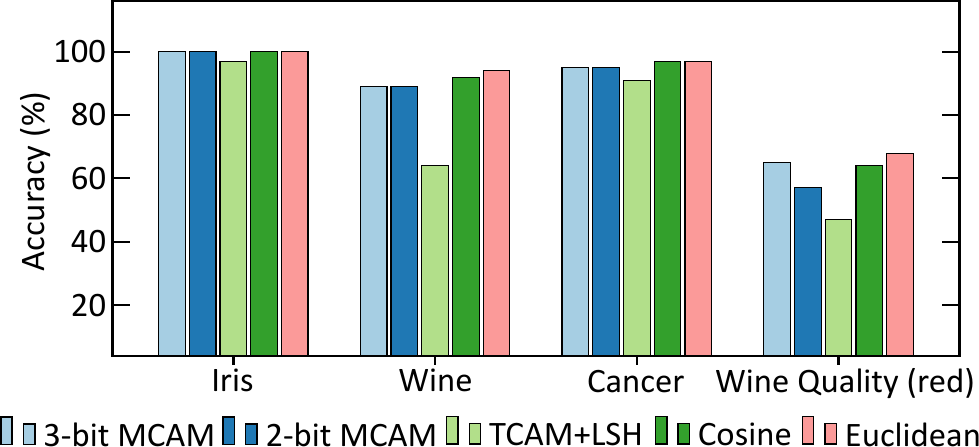}
  \caption{Accuracy of NN classification tasks for different distance functions.}
  \label{fig:classification}
  \vspace{-0.2in}
\end{figure}

NN classification is a conventional, non-parametric method to classify data which is commonly used as the baseline method in many pattern classification problems~\cite{hu2016distance}. To benchmark the MCAM distance function for NN classification tasks, we consider the top 4 most cited datasets in the UCI ML repository~\cite{Dua2019} that only contain real-valued, non-categorical data, namely, Iris, Wine, Breast Cancer, and Wine Quality. We randomly split each dataset into training (80\%) and test (20\%) sets. Given a query from the test set, we search for the sample in the training set with the smallest distance given a distance function, the label associated with the best match is returned as the label for the query. For the MCAM and TCAM results, we assume that the CAM words have the same number of cells as the number of features in the dataset.

Fig.~\ref{fig:classification} shows the results of NN classification evaluations. The 2-bit and 3-bit MCAM implementations leveraging the proposed distance function consistently outperform the TCAM+LSH while achieving accuracies comparable to software implementations (Cosine and Euclidean). The 3-bit MCAM achieves 12\% higher accuracies on average compared to TCAM+LSH. The 2-bit MCAM performs on par or slightly worse than the 3-bit MCAM for all the datasets. Although we generally expect the 3-bit MCAM to outperform the 2-bit MCAM due to the higher precision, simpler tasks such as NN classification do not benefit from that extra precision. 

\subsection{One/Few-Shot Learning}
\label{sec:few-shot}
One/few-shot learning aims to classify unseen images with only one/few images for training~\cite{vinyals2016matching}. MANNs are extremely attractive for one/few-shot learning applications. MANNs are comprised of a neural network for feature extraction and a memory module for storing and loading features where the features are the outputs of the last layer of the neural network. The memory module holds the features of trained classes which can be used to classify previously unseen images. To perform inference with MANNs, the features of the query image are extracted using the neural network and compared with the features of the trained classes stored in memory. The label of the nearest neighbor to the query $q$ among the $M$ classes is chosen as the label of $q$.

The MANN used for these tasks follows the implementation in~\cite{wang2019simpleshot} which achieves state-of-the-art accuracies. The neural network part of the MANN is comprised of two 3$\times$3 convolution layers with 64 filters, a max-pooling layer, two 3$\times$3 convolution layers with 128 filters, and a max-pooling layer followed by two 128 and 64 node fully-connected layers. Since there are 64 nodes in the last fully-connected layer of the neural network, the queries have 64 features. Thus, we assume that the TCAM and MCAM have 64 cell words.

\begin{figure}[t]
  \centering
  \includegraphics[width=0.93\linewidth]{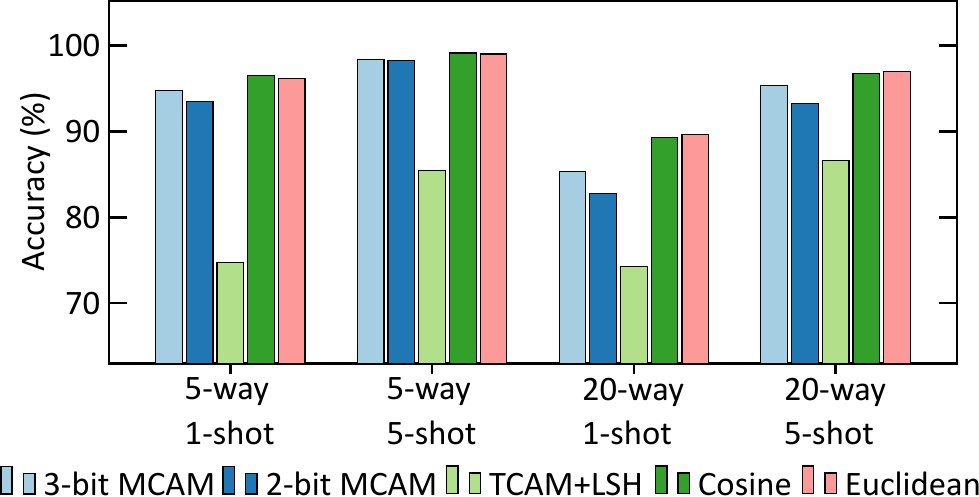}
  \caption{Accuracy of few-shot learning tasks for different distance functions.}
  \label{fig:mann_acc}
  \vspace{-0.2in}
\end{figure}

Fig.~\ref{fig:mann_acc} shows the accuracy of the one/few-shot learning tasks for the Omniglot dataset~\cite{lake2015human}. For a $N$-$way$ $K$-$shot$ task, the network trains on $N\times K$ images for $K$ classes ($N$ images per class). The 2-bit and 3-bit MCAMs on average outperform TCAM+LSH by 11.6\% and 13\%, respectively, while achieving accuracies comparable to the FP32 cosine and Euclidean. As an example, for the 5-way 1-shot task, the MCAM achieves 98.34\% accuracy (only 0.8\% and 0.7\% lower than cosine and Euclidean, respectively). This is a 13\% improvement over TCAM+LSH which is significant\footnote{The TCAM+LSH results presented in~\cite{ni2019ferroelectric} are higher than what we report because they use 512-bit LSH signatures that require 512-bit TCAM words.}.


The proposed distance function is as effective for NN search applications as the FP32 software implementations. This shows the efficacy of the proposed distance function even at low-bit precisions such as 2 and 3 bits. Significant accuracy improvements of the 2-bit and 3-bit MCAMs over the TCAM+LSH approach are mainly due to the fact that the MCAM allows an \textit{exact} NN search with the proposed distance function, whereas TCAM+LSH tries to \textit{approximate} the cosine distance function for NN search. Note that the MCAMs achieve competitive accuracies compared to \textit{exact} NN search based on cosine distance function. Lastly, with CAMs of same word length, the MCAM cells enable a higher bit representation for the features.

\begin{figure}[t]
  \centering
  \includegraphics[width=0.95\linewidth]{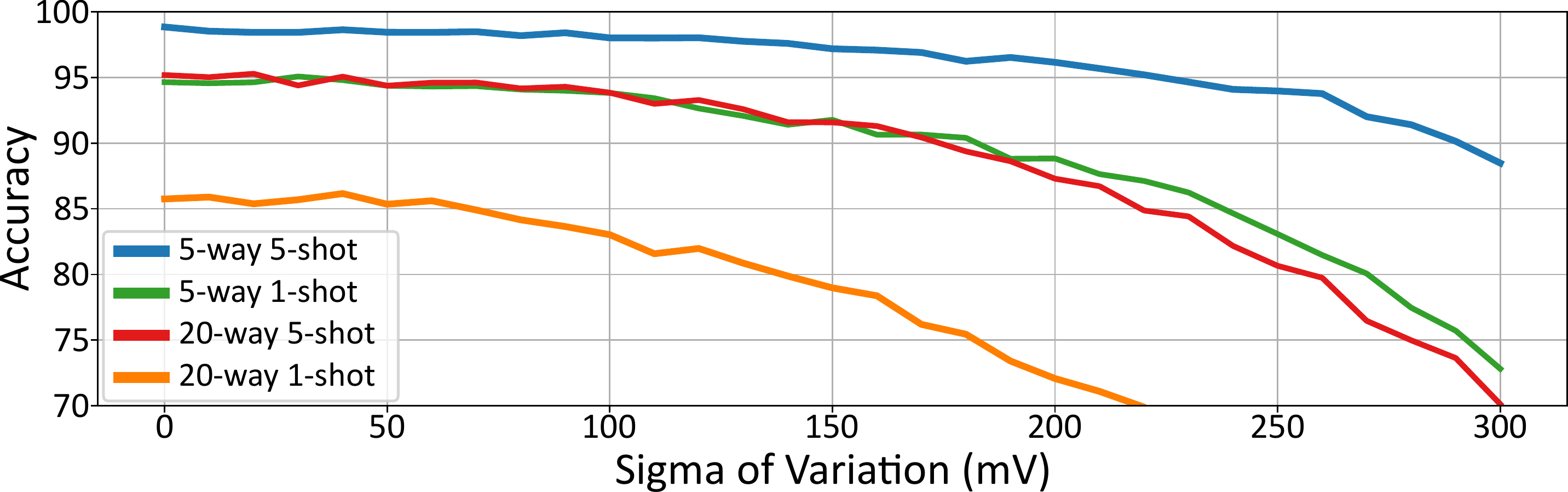}
  \caption{One/few-shot learning results for a 3-bit MCAM with $V_{th}$ variations.}
  \label{fig:varitation_acc_results}
  \vspace{-0.215in}
\end{figure}

Fig.~\ref{fig:varitation_acc_results} shows the accuracy of one/few-shot learning tasks for a 3-bit MCAM with respect to sigma of the FeFET $V_{th}$ distributions. It is notable that results do not suffer any accuracy loss for sigma values of up to 80mV which is the highest observed in our simulations in Sec.~\ref{sec:vth_var}. These results suggest that the proposed distance function has high $V_{th}$ variation tolerance for one/few-shot learning applications. This is quite favorable for FeFET MCAM design as current FeFET devices in the literature exhibit $V_{th}$ variations when programmed without any verify pulses.

We evaluate the energy and delay of MCAMs for one/few-shot learning under the same set of assumptions in~\cite{ni2019ferroelectric}. Since the TCAM and MCAM cells are the same, use the same sensing scheme, and use programming pulses of the same width, same-sized MCAMs and TCAMs have the same search and programming delay. Moreover, average programming energy of the MCAM is 12\% lower than the TCAM, due to lower programming voltages. However, the average energy of search is 56\% higher for the MCAM due to higher search voltages. Following the distribution in~\cite{ni2019ferroelectric}, both TCAM and MCAM offer end-to-end improvements of $4.4\times$ and $4.5\times$ in terms of energy and latency, respectively, compared to a Jetson TX2 GPU implementation (same as in~\cite{ni2019ferroelectric}) for one/few-shot learning. Although energy improvements of MCAMs and TCAMs are not the same, the end-to-end improvements for this application are bound by the neural network part of the MANN. Thus, improvement numbers are simliar to~\cite{ni2019ferroelectric}.

\subsection{Experimental Demonstration of FeFET MCAM}

To further demonstrate the feasibility of the FeFET MCAM, we use FeFETs manufactured and embedded in a 28-nm high-k metal gate technology by \mbox{GLOBALFOUNDRIES}~\cite{trentzsch201628nm}. The transistors have a channel length and width of 450nm and 450nm. The FeFETs are arranged in an AND array structure~\cite{cagli2019performance}. The equivalent MCAM cell (Fig.~\ref{fig:mcam_schematic_and_vbs}(a)) is constructed by using two FeFETs that are connected along their drain contacts while source contacts are connected to ground. The FeFETs are set to the corresponding $V_{th}$ states using a single same-width pulse scheme with varying amplitude. For setting the high $V_{th}$ state the FeFET is erased using a gate voltage of -5V with a pulse time of 500ns. The applicable voltage range for programming the intermediate $V_{th}$ states is 1V to 4.5V in steps of 0.1V with pulse times of 200ns. For a 2-bit demonstration, four evenly distributed target $V_{th}$ values are defined. After setting the FeFET states, MCAM conductance is obtained by setting ML to 0.1V and by measuring the ML current over a DL sweep from -0.5V to 1.1V.

\begin{figure}[t]
  \centering
  \includegraphics[width=\linewidth]{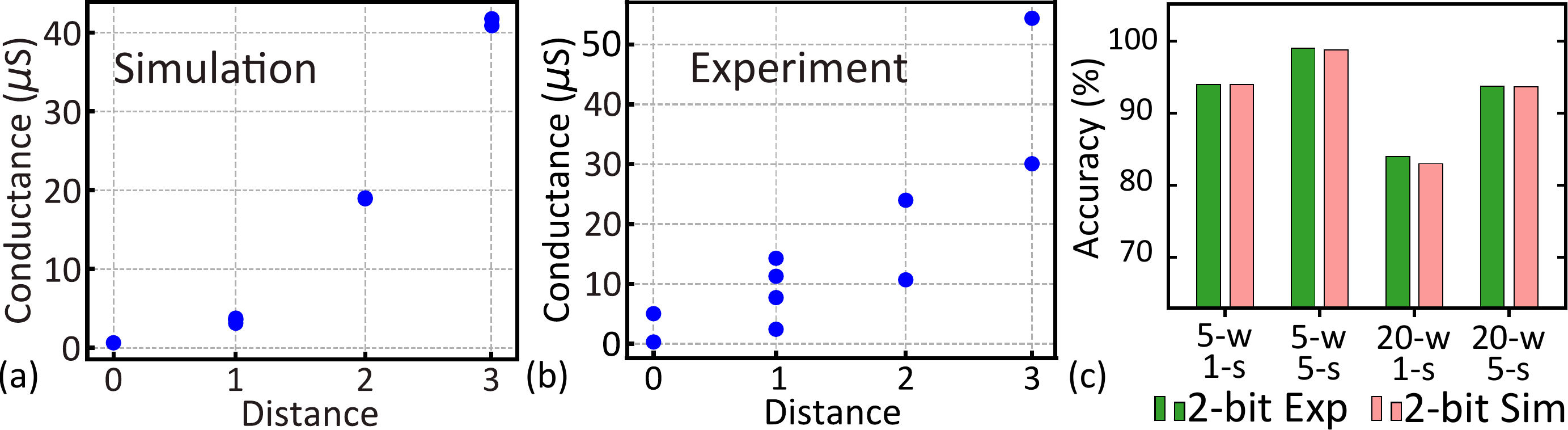}
  \caption{(a) and (b) depict the distance function of a 2-bit FeFET MCAM in simulation and in experiment, respectively. (c) one/few-shot learning results.}
  \label{fig:experimental}
  \vspace{-0.21in}
\end{figure}

Fig.~\ref{fig:experimental}(a) and Fig.~\ref{fig:experimental}(b) show the distance function of a 2-bit FeFET MCAM based on simulation and experiment, respectively. Experimental results follow the trends of simulations where the conductance increases exponentially with higher distances. Note that these results are preliminary efforts and we aim to demonstrate 3-bit precision in the future. There are different techniques that allow for better control over the FeFET polarization switching such as write-and-verify which can be explored for further improvements. Fig.~\ref{fig:experimental}(c) shows the results of one/few-shot learning with the 2-bit experimental data in Fig.~\ref{fig:experimental}(b). Results show that even with the current experimental data, acceptable application-level accuracies can be achieved. We even observe a higher accuracy for the experimental distance function due to its noisy behavior which helps with regularization~\cite{li2018efficient}. This is in-line with our findings on FeFET $V_{th}$ variations which mimics a noisy behavior.

\section{Conclusion}
\label{sec:conclusion}
We proposed a novel distance function for NN search via FeFET MCAMs. The proposed distance function: (i) improved the accuracy of one/few-shot learning tasks for the Omniglot dataset by 13\% on average with iso-energy and delay compared to the approach in~\cite{ni2019ferroelectric}, (ii) achieved software-equivalent accuracies for NN classification tasks, (iii) was resilient to FeFET $V_{th}$ variations for one/few-shot learning, and (iv) was experimentally demonstrated with a 2-bit FeFET MCAM. Finally, our analysis of FeFET MCAM distance function is applicable to other MCAMs as well.

\section*{Acknowledgment}
This work was supported in part by ASCENT, one of six centers in JUMP, a SRC program sponsored by DARPA. We received funding within ECSEL Joint Undertaking in collaboration with the European Union’s H2020 Framework Program and National Authorities, under grant agreement number 826655. We thank GLOBALFOUNDRIES for the provision of 28nm technology FeFET structures.

{\footnotesize
\bibliography{bibfile}}
\bibliographystyle{./my_abbrv.bst}

\end{document}